\documentclass[12pt,preprint]{aastex}
\shorttitle{formation of isothermal CPDs}
\shortauthors{Wang et al.}
\usepackage{bm}
\usepackage{sansmath}
\usepackage[usenames,dvipsnames]{color}
\newcommand{\hhwang}[1]{{\color{black} #1}}

\newcommand{\wang}[1]{{\color{black} #1}}

\begin{document}

\title{Formation of Isothermal Disks around Protoplanets. I. Introductory Three-Dimensional Global Simulations for Sub-Neptune-Mass Protoplanets}

\author{Hsiang-Hsu Wang}
\affil{Institute of Astronomy and Astrophysics, Academia Sinica, P.O. Box 23-141, Taipei 10617, Taiwan, R.O.C.}
\email{hhwang@asiaa.sinica.edu.tw}

\author{Defu Bu}
\affil{Key Laboratory for Research in Galaxies and Cosmology, Shanghai Astronomical Observatories, Chinese Academy of Sciences, 80 Nandan Road, Shanghai 200030, China.}
\affil{Key Laboratory of Planetary Sciences, Chinese Academy of Sciences, Shanghai ,200030, China}

\and

\author{Hsien Shang and Pin-Gao Gu}
\affil{Institute of Astronomy and Astrophysics, Academia Sinica, P.O. Box 23-141, Taipei 10617, Taiwan, R.O.C.}

\begin{abstract}
\hhwang{The regular satellites found around Neptune ($\approx 17~M_{\Earth}$) and Uranus ($\approx 14.5~M_{\Earth}$) suggest that past gaseous circumplanetary disks may have co-existed with solids around sub-Neptune-mass protoplanets ($< 17~M_{\Earth}$). These disks have been shown to be cool, optically thin, quiescent, with low surface density and low viscosity. Numerical studies of the formation are difficult and technically challenging. As an introductory attempt, three-dimensional global simulations are performed to explore the formation of circumplanetary disks around sub-Neptune-mass protoplanets embedded within an isothermal protoplanetary disk at the inviscid limit of the fluid in the absence of self-gravity. Under such conditions, a sub-Neptune-mass protoplanet can reasonably have a rotationally supported circumplanetary disk. The size of the circumplanetary disk is found to be roughly one-tenth of the corresponding Hill radius, which is consistent with the orbital radii of irregular satellites found for Uranus. The protoplanetary gas accretes onto the circumplanetary disk vertically from high altitude and returns to the protoplanetary disk again near the midplane. This implies an open system in which the circumplanetary disk constantly exchanges angular momentum and material with its surrounding prenatal protoplanetary gas.} 
\end{abstract}


\section{Introduction}
\hhwang{Regular satellites of giant planets in the Solar System are commonly postulated to form in circumplanetary disks (CPDs, hereafter) surrounding their parent planets. The process of forming a CPD accumulates gas and rock-ice solids, which are thought to be raw material of satellites, from the parent protoplanetary disk.  Properties of regular satellites, such as their chemical compositions, differentiation, total mass, as observed \hhwang{for} the outer planets (Jupiter, Saturn, Uranus, Neptune) in the Solar System, place tight constraints on their precursors. Icy satellites would not have formed if the CPDs formed were too hot for the volatiles to condense. Very dense CPDs would not have enough time to allow satellites to accrete mass and to survive through fast orbital decays. Understanding the CPDs is a key step toward understanding the formation of satellites.}

\hhwang{Several models for CPDs have been proposed to account for the formation of satellites systems around giant planets. The solid enhanced minimum mass (SEMM) disk model \citep{Mos03a,Mos03b} suggests that CPDs formed around gas giants (Jupiter and Saturn) may be optically thick inside and optically thin outside, while those formed around ice giants (Uranus and Neptune) may be isothermal to the background nebula throughout the disk due to their low surface density. While the inner Saturnian satellites can only acquire volatiles after the disk cooled down, Uranian/Neptunian satellites may accrete (icy) mass immediately after the formation of Uranus/Neptune. The CPD formed around Uranus/Neptune is supposed to be cool, quiescent, and laminar, with low viscosity. The low-surface density also suggests that the self-gravity of gas in those isothermal CPDs can be neglected as a first approximation (see \citet{Mos10} for the discussions on the Jovian and Saturnian systems). \citet{Can02,Can06} proposed an alternative ``gas-starved" CPD produced by a slow inflow of gas and solids. Similar to SEMM disks, such CPDs are expected to be optically thin and laminar, and with low viscosity, expected for the formation of Galilean satellites. The major difference between the two competing models is how a CPD accretes gas and solids. More recently, a gas-free tidal-spread model is advocated \citep{Cri12} to explain the puzzling mass-distance trend as observed in the outer three giant planets. However, this picture is hard to reconcile with the formation of Galilean satellites, which may still need to be assembled in a gaseous environment.}

\hhwang{The formation of CPDs is complex and is best studied with numerical simulations \citep{Lub99,Dan02,Tan02,Dan03,Bat03,Mac08, Mac09, Ayl09b,Mac10,War10,Mar11,Tan12,Ayl12,Uri13}. It is especially technically challenging to simulate the formation of CPDs surrounding low-mass protoplanets like Uranus and Neptune. For example, for a \hhwang{17~$M_{\Earth}$} protoplanet embedded within a protoplanetary disk with an aspect ratio 0.05 at the heliocentric distance 5.2~AU, the corresponding Hill radius is only half of the scale height of the protoplanetary disk. The CPD formed around is deeply embedded inside the scale height of the disk. The gas motions are three-dimensional in the surroundings of low-mass protoplanets. These CPDs are not isolated objects attached only to the protoplanets. They are constantly exchanging material and get disturbed from the bigger protoplanetary disks. For motions in and out of the CPDs in such environment, the inclusion of geometric curvature is also necessary. Three-dimensional global simulations are required to cover the system. Using high spatial resolution without introducing softening lengths is the key to correctly follow the gasdynamics of CPDs without distorting gas behaviour around protoplanets for a CPD whose size is small compared to the corresponding Hill sphere \citep{Qui98,Mar11}.  }

\hhwang{Due to the technical challenges, only a handful of numerical investigations have been attempted for the CPDs around low-mass protoplanets. \citet{Bat03} performed three-dimensional global simulations to explore formation of CPDs around protoplanets of one Earth mass to one Jupiter mass embedded in locally isothermal protoplanetary disks. Owing to insufficient spatial resolution ($\approx$0.015~AU), CPDs are resolved only for protoplanetary masses greater than 32~$M_{\Earth}$, making the formation of CPDs around protoplanets of mass less than 32~$M_{\Earth}$ inconclusive. \citet{Ayl09a} performed sophisticated three-dimensional local smooth particle hydrodynamics (SPH) simulations including self-gravity of gas and interstellar grain opacities to study \hhwang{accretion of gas onto prescribed solid cores}. \citet{Ayl09b} found no CPDs can form around sub-Neptune-mass protoplanets even with a locally isothermal equation of state. The conclusion obtained from those numerical experiments, however, poses a challenge to current understanding of satellite formation. \citet{Cri12} proposed that regular satellites of Uranus and Neptune formed from ancient massive rings rather than in a CPD, as partly motivated by \citet{Ayl09b}. }

\hhwang{Improvements that address the challenges are needed in state-of-the-art numerical codes to attack the physical problems. Several new features have been implemented into our Antares code \citep{Yua05}. The original two-dimensional Cartesian code is extended to solve the three-dimensional hydrodynamic equations in cylindrical coordinates so that the effects of the third dimension and geometric curvature are included in this setup. Global simulations are performed to avoid the use of artifical boundary condition in the azimuthal direction. Nested-Mesh-Refinement (NMR) is also implemented to resolve both the protoplanetary disk ($\approx 10$~AU) and CPDs ($\approx 0.01$~AU) that are of different dynamical ranges, and enables us to concentrate computational power on the surroundings of protoplanets. The use of softening length is avoided in this work not  to distort the potential of protoplanets. This allows us to properly follow the gas behaviour of CPDs. The global disk is relaxed for 8 orbital times measured at the location of protoplanets for a carefully prepared initial condition. The perturbation in the protoplanetary gas caused by low-mass protoplanets is relatively small and can easily be overwhelmed without being started in an initial state that is as close to equilibrium as possible.}

\hhwang{We address the formation of CPDs around sub-Neptune-mass protoplanets under the context of satellite formation using three-dimensional global hydrodynamic simulations.} This work is parallel to the locally isothermal work of \citet{Ayl09a} but is performed with a different numerical scheme. This paper is structured as follows. The physical models and a brief introduction of the numerical method are described in \S~2. \hhwang{Analytic expectations based on a simple assumption and results obtained from numerical simulations are presented and analysed in \S~3. A brief summary, discussions and implications of our findings are arranged in \S~4.}

\section{Models}
\subsection{Governing equations}
We explore the formation of CPDs formed around sub-Neptune-mass protoplanets with masses of 4, 8 and 16~$M_{\Earth}$. Since the Hill radii of these models are smaller than the scale height of the protoplanetary disk, three-dimensional models are important for correctly following the flow patterns around protoplanets. The governing equations are described using cylindrical coordinates $(r,\phi,z)$ co-rotating with protoplanets and with the origin centered at a host star of one solar mass. Here, we ignore the small shift of position between the central star and the center of mass. The globally isothermal gaseous disk then evolves based on the following governing equations without considering the self-gravity and the viscosity of gas \citep{Dan02,Ski10}:

\begin{eqnarray}
\frac{\partial \rho}{\partial t}+\frac{1}{r}\frac{\partial}{\partial r}(r\rho v_r)+\frac{1}{r}\frac{\partial}{\partial \phi}(\rho v_{\phi})+\frac{\partial}{\partial z}(\rho v_z)=0, \\
\frac{\partial}{\partial t}(\rho v_r)+\frac{1}{r}\frac{\partial}{\partial r}\left[r\left(\rho v_r^2+p \right)\right]+\frac{1}{r}\frac{\partial}{\partial \phi}\left(\rho v_r v_{\phi}\right)+\frac{\partial}{\partial z}\left(\rho v_r v_z\right)=\left( \rho v_{\phi}^2+p\right)/r+\rho\left(2\Omega_p v_{\phi}+\Omega^2_p r -\frac{\partial \Phi}{\partial r}\right), \\
\frac{\partial}{\partial t}\left(\rho v_{\phi}\right)+\frac{1}{r}\frac{\partial}{\partial r}\left(r\rho v_r v_{\phi}\right)+\frac{1}{r}\frac{\partial}{\partial \phi}\left(\rho v^2_{\phi}+p\right)+\frac{\partial}{\partial z}\left( \rho v_{\phi} v_z\right)=-\left(\rho v_r v_{\phi}\right)/r+\rho\left(-2\Omega_p v_r-\frac{1}{r}\frac{\partial \Phi}{\partial \phi}\right),\\
\frac{\partial}{\partial t} \left(\rho v_z\right)+\frac{1}{r}\frac{\partial}{\partial r}\left(r\rho v_r v_z\right)+\frac{1}{r}\frac{\partial}{\partial \phi}\left(\rho v_z v_{\phi}\right)+\frac{\partial}{\partial z}\left(\rho v_z^2+p\right)=-\rho\frac{\partial \Phi}{\partial z},
\end{eqnarray}
with $\rho$ being the volume density, $(v_r,v_{\phi},v_z)$ the velocity components in the radial, azimuthal and vertical directions observed in the rotating frame, respectively, $p=c^2_s \rho$ the isothermal gas pressure with $c_s$ the sound speed, $\Omega_p$ the orbital angular speed of the protoplanet, $\Phi$ the total gravitational potential contributed from the central star and the protoplanet. 

Equation (1) is the continuity equation, Equations (2) to (4) are  the momentum equations in the radial, azimuthal and vertical directions, respectively. Notice that the left-hand side of Equations (1) to (4) is arranged in a conservation form, which allows us to make use of the original exact Riemann solver used for the Cartesian Antares with minimum modification \citep{Yua05}. The first terms on the right-hand side of Equations (2) and (3) are geometric source terms. The Coriolis force, the fictitious centrifugal force due to the non-inertial frame and the gravitational forces contributed from the central star and the protoplanet are organized together in the right-most of Equations (2) to (4). The gravitational potential $\Phi$ has two parts given by
\begin{equation}
\Phi=-\frac{G {M}_{\odot}}{\sqrt{r^2+z^2}}-\frac{G {M}_p}{\sqrt{R^2+z^2}},
\end{equation}
where $M_{\odot}$ is one solar mass, $M_p$ denotes the mass of protoplanet, $G$ the gravitational constant, $r$ and $R$ are the radial distances measured from the central star and from the protoplanet, respectively. We note that the softening length is not introduced explicitly in our models. $R$ is related to the coordinates $(r, \phi, z)$ through the following relation:
\begin{equation}
R^2 = r^2+r_p^2-2rr_p\cos(\phi),
\end{equation}
where $r_p$ is the orbital radius of the protoplanet, which is fixed at 5.2~AU for all models in this work. The isothermal sound speed, $c_s$, is chosen through the following relation:
\begin{equation}
\frac{c_s}{v_{\rm Kep}}=\frac{H}{r_p}=0.05,
\end{equation}
with $H$ being the disk thickness at $r_p$ and $v_{\rm Kep}=\sqrt{G{\rm M}_{\odot}/r_p}=\Omega_p r_p$ the corresponding Keplerian orbital velocity. \hhwang{The corresponding temperature at this particular orbital radius, $r_p$, can be estimated with a standard model of solar minimal nebula \citep{Hay81,Hay85} through the relation:

\begin{equation}
T=280.0\left(\frac{L}{L_{\odot}}\right)^{1/4}\left(\frac{r_p}{1{\rm AU}}\right)^{-1/2}~{\rm \wang{K}},
\end{equation}

where $L$ and $L_{\odot}$ are the protostellar and solar luminosity, respectively. The gas temperature at the orbital radius is then evaluated to be $T=123$~K if $L=L_{\odot}$ is adopted.} Since we are interested in the local phenomena in the neighbourhood of protoplanets, given also the possibility that CPDs formed around sub-Neptune-mass planets are optically thin \citep{Mos03a}, we further assume the temperature is uniform in the whole computational domain. \hhwang{It should be noted here that, in general, a global simulation using constant temperature is not an appropriate assumption. However, since the circumplanetary disks formed around low-mass protoplanets only accrete surrounding gas of nearly the same temperature and the orbits of protoplanets are fixed at a constant radius, we expect that global isothermal assumption adopted in this particular work would not dramatically affect our general conclusions for the properties of CPDs.} The purpose of performing global simulations is to provide an appropriate background flow and to avoid using artificially imposed boundary conditions in the azimuthal direction.

\subsection{Computational setup}
Three-dimensional global isothermal simulations are performed to study the formation of CPDs around sub-Neptune-mass protoplanets. Since the simulations are evolved in a frame co-moving with protoplanets, we fix the location of protoplanets at $(r_p,\phi_p,z_p)=(5.2,0,0)$AU. The choice of placing protoplanets at 5.2AU is mainly for the comparison between our conclusions and that of \citet{Ayl09b}, though also supported with the model proposed by \citet{Tho02}. The protoplanetary disk is modeled in the region $r\in [2, 8]$AU, $\phi\in [0, 2\pi]$ radian and $z\in [0, 1]$AU. The root grid (coarsest grid, zeroth level)  is covered by $200 \times816 \times25$ cells uniformly distributed in the radial, azimuthal and vertical directions, respectively. The cell numbers are chosen such that the shape of cells close to a protoplanet is nearly a cube. The protoplanet is placed in the corner of cells to avoid singularity of gravitational potential. We note that no protoplanet surface is modeled and therefore any effects related to the boundary layer around protoplanets are not studied in this work. \hhwang{A possible impact of the presence of boundary layer is the release of gravitational energy into the surroundings as a source of heat when the cores still accrete planetesimals. Since those CPDs of our interest are at the late stage of planet formation and are inviscid, we assume that the accretion onto core surfaces is almost subsided, though satellitesimals may still fall in due to gas drag or to tidal torques \citep{Can06}. Furthermore, if we simply adopt the mean density estimated for the ice-rock core of Uranus \citep{Pod95} and apply it to evaluate the core radius of our models (see Tabel~\ref{table:properties}), one will find that the numerical resolution near the protoplanets are about 4 to 7 times these core sizes. The physics near the `surface' of solid cores are not resolved in this work and therefore are neglected. \wang{The outer gaseous envelope, which is not considered in the estimates of ice-rock core sizes, contributes roughly additional 20\% to 30\% radial extension in the sizes based on the models constructed for the internal structures of Uranus and Neptune \citep{Pod95}. We note that at the epoch when protoplanets just formed, the entropy in the gaseous envelopes may be high, so that the envelopes could be more extended than what are presently observed.} }

Since the dynamical range involved in the formation of a CPD spans three orders of magnitude from 10AU to 10$^{-2}$AU, we adopt the numerical technique so-called Nested-Mesh-Refinement (NMR) to evolve both the protoplanetary and circumplanetary disks. With this technique, spatial resolution uniformly increases in a nested fashion, i.e., a uniform mesh of higher spatial resolution of level $l$ is embedded within a uniform coarse mesh of level $l-1$. In this way, computational power can be concentrated around the areas of scientific interests, i.e., the CPDs in our work.  The volume covered by higher level grids follows the rules:
\begin{eqnarray}
|r-r_p|_l=\frac{3}{2^l},\\
|\phi-\phi_p|_l=\frac{\pi}{2\cdot 2^l},\\
|z-z_p|=\frac{1}{2^l},
\end{eqnarray}
where $l$ is an integer running from 1 to 6, representing the level of refinements. The spatial resolution of level $l$ doubles that of the level $l-1$. With this rule, the finest spatial resolution is $6.25 \times 10^{-4}$AU (cf. Jupiter's radius $4.77\times 10^{-4}$AU). \hhwang{In terms of the finest spatial resolution}, the corresponding Hill radii for protoplanets with 4, 8 and 16M$_{\Earth}$ located at 5.2AU will be \hhwang{resolved} with cell numbers \hhwang{$N_{R_H}=134, 168$ and $212$}, respectively. Since the frame is co-rotating with protoplanets, the grid structure is static in time. Because the equations we solve for are symmetric with respect to the midplane, only the \hhwang{north hemisphere} is considered. The initial disk density profile was chosen to be axisymmetric and to follow:
\begin{equation}
\rho(r,\phi,z) \propto \frac{1}{r^2}\exp\left(-\frac{G{\rm M}_{\odot} z^2}{2r^3c_s^2}\right),
\end{equation}
where the volume density at the disk midplane scaling as $r^{-2}$ is an arbitrary choice. Using different power laws are expected not to affect the results significantly because of the small size of a CPD compared to the characteristic length of the protoplanetary disk. The volume density is scale free because the self-gravity of gas is neglected in this work. The initial velocity $(v_r,v_{\phi},v_z)$, in the co-rotating frame reads:
\begin{equation}
(v_r,v_{\phi}, v_z)=(0,v_{\rm Kep}-\Omega_p r, 0), 
\end{equation}
where $v_{\rm Kep}$ denotes the circular Keplerian velocity orbiting the central star for a given radius. The physical boundary conditions are all fixed using the initial condition described above except that the boundary at the midplane is reflective. \hhwang{Note that no boundary in the azimuthal direction needs to be specified in global simulations}.

\subsection{Numerical method}
Three dimensional simulations are performed with a high-order Godunov code known as Antares, in which hydrodynamic fluxes on cell interfaces are obtained from the exact/approximate Riemann solution \citep{Yua05}. We employ the finite volume method to solve the hydrodynamic equations outlined in Sec.~2.1. The nested grid used in this work is based on the AMR engine implemented for resolving a huge dynamical range. For the current work, computational domains are refined according to Eqs. (9) to (11) only at the beginning of a simulation. The grid structure is then fixed without change. 

The grid arrangement of levels looks the same as shown in the Fig.~1 of \citet{Dan02}. Unlike the staggered grid used in their work, we put all the variables at cell centers. The second-order accuracy in space is achieved by utilizing slope-limiters, while the second-order accuracy in time is implemented with the Runge-Kutta method of second order (RK2). A global time step, $\Delta t$, is determined by fulfilling the Courant-Friedrichs-Lewy (CFL) condition for all grids of different levels. The boundary of each grid is surrounded by two layers of ghost cells. The variables of ghost cells which are adjacent to the physical boundaries are described by Eqs. (12) and (13). For the ghost cells between levels $l$ and $l+1$, variables of ghost cells on level $l+1$ are evaluated by conservative interpolations between the associated cells of level $l$ \citep{Li04}. The flux correction and the variable restriction are forced between levels to make sure the conservation of conservative variables between levels. 

The exact Riemann solver is used for the calculation of hydrodynamic fluxes. We found that the exact Riemann solver is more robust than the approximate Riemann solver of HLL-type when the softening length is not explicitly introduced. In the Antares code, the Riemann problem is first solved in an iterative way. If failed, the bisection method is followed to find the solution. 

\subsection{Preparation for the initial condition}
A good initial condition is especially important for the growth of CPDs around sub-Neptune-mass planets. The energy involved in the relaxation process purely from the numerical discretization may be comparable to the potential energy of low-mass protoplanets. This may potentially bias our conclusions, making the evolution of the first few orbits untrustworthy. To get around this problem, we first relax the initial condition as described in Sec.~2.2 for 8 orbital times (measured at the position of protoplanets, i.e., 5.2 AU) until the protoplanetary disk accommodates itself to the nested grid structures. After the relaxation, we expect the new configuration should not deviate much from the imposed initial condition. 

\begin{figure}
\epsscale{1.1}
\plotone{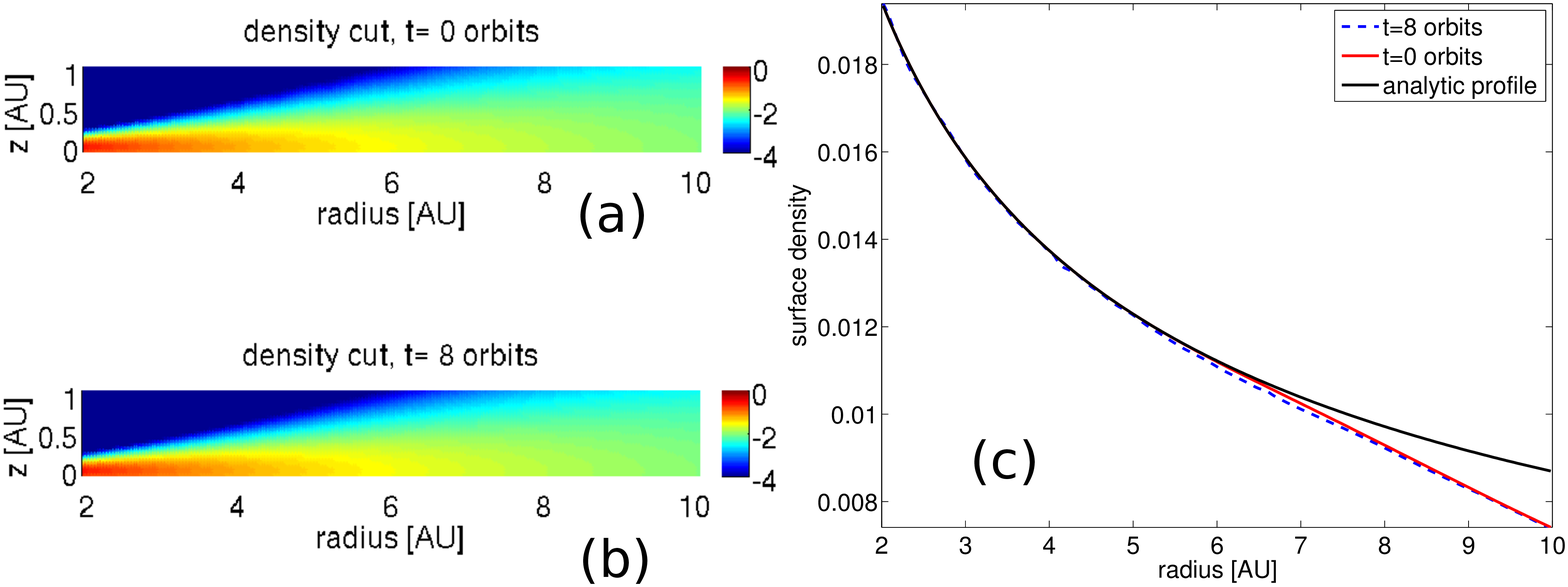}
\caption{Initial condition prepared for three-dimensional simulations. Before the planetary potential is turned on, the imposed basic state is relaxed for 8 orbital times. Images (a) and (b) show the volume density (on logarithmic scale) cutting through the plane $\phi=0$ at $t=0$ and $t=8$ orbits, respectively. (c) The vertically integrated surface density along the line $\phi=0$ after the relaxtion (dashed line) is almost identical to that of the imposed initial condition (red solid). \hhwang{The analytic profile (black solid) is obtained by direct integrating Eq.~(12) for a given midplane density from $z=0$ to $\infty$. The good match inside 6AU indicates that the vertical structure of the disk is well-resolved with our coarsest numerical resolution, while the deviation in the outer disk reflects that the computational domain is not large enough to include all disk gas.} Since the self-gravity of gas is not taken into account, both \hhwang{the} volume density and 
\hhwang{the} surface density are scale free. \label{fig:IC}}
\end{figure}

Figure~\ref{fig:IC} illustrates the result after a relaxation of 8 orbital times. Figures~\ref{fig:IC}a and \ref{fig:IC}b show the volume density cutting through the plane defined by $\phi=0$ (on logarithmic scale) at $t=0$ and $t=8$ orbits, respectively. Figure~\ref{fig:IC}c shows the vertically integrated surface density along $\phi=0$.  The surface density after relaxation (dashed line) is almost identical to the imposed initial condition (red solid). We adopt the relaxed configuration as our {\it real} initial condition throughout this work. The potentials of protoplanets are gradually turned on and reach their full strength after 0.4 orbital time.  \hhwang{The analytic radial profile (black solid) of the surface density integrated directly by Eq.~(12) from $z=0$ to $\infty$ is also shown in Fig.~\ref{fig:IC}c for comparison. The good match between the analytic profile and the numerical setup inside 6AU indicates that the vertical structure of the protoplanetary disk is well-resolved with the coarsest numerical resolution, whereas the deviation from analytical profile in the outer disk reflects that the computational domain is not large enough to include the high altitude disk gas. Since the circumplanetary disks around low-mass protoplanets only accrete material from the nearby surroundings and the Hill radii of our models are less than the scale height of the protoplanetary disk, we expect that missing high altitude gas in the outer disk would not alter our general conclusions.} 

\section{Results}
\begin{figure}
\epsscale{1.1}
\plotone{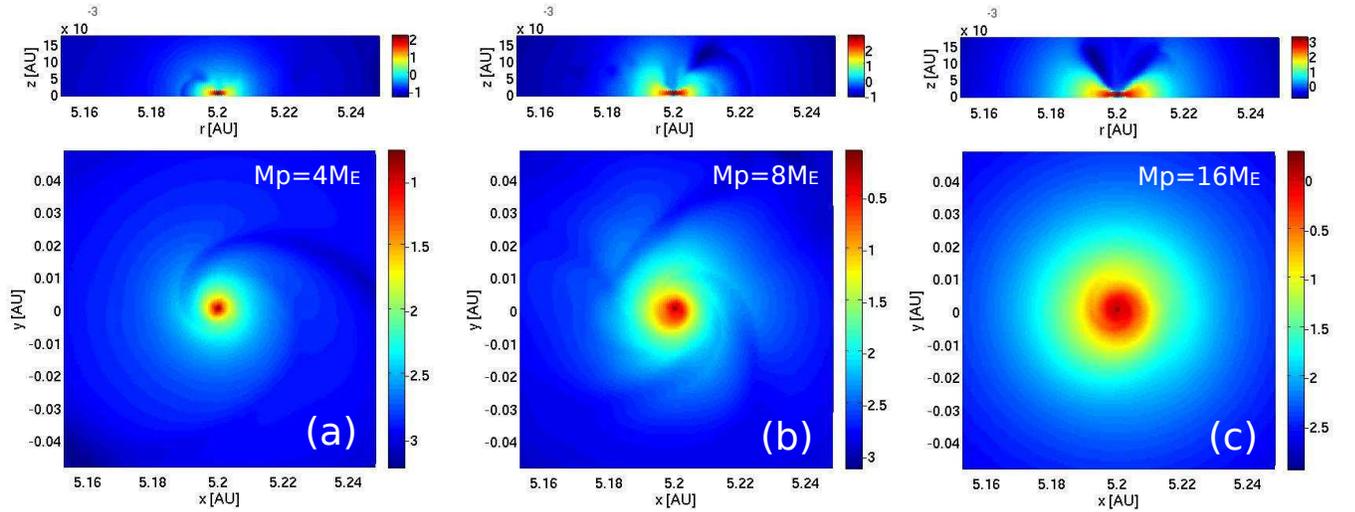}
\caption{\hhwang{The bottom panel shows square sections of surface density centered at the locations of protoplanets. The surface density is obtained by vertical integtation over $z<0.0156$AU (the height of the finest level) for protoplanets with masses 4, 8 and 16M$_{\Earth}$, respectively. The central star is located on the left and the protoplanets are orbiting the central star counterclockwise.} On top of each surface density is the corresponding volume density cutting through the plane defined by $\phi=0$.  Since the self-gravity of gas is not taken into account, both \hhwang{the} volume density and \hhwang{the} surface density are scale free. \label{fig:edge_face_view}}
\end{figure}

The scientific goal of this work is to show that the expected CPDs, \hhwang{which is associated with the formation of satellite}, can \hhwang{reasonably} form around sub-Neptune-mass protoplanets in an isothermal, \hhwang{inviscid and non-self-gravitating} protoplanetary disk. The upper panel of Fig.~\ref{fig:edge_face_view} shows the edge-on view of volume density cutting through the plane defined by $\phi=0$. The bottom panel shows the corresponding surface density integrated vertically for the finest grid, i.e., \hhwang{from $z=0$ to $0.0156$}AU as shown in the figures. \hhwang{It is clear that the disk structure is increasingly evident with increasing protoplanetary mass due to the density contrast between CPDs and high altitude protoplanetary gas. That is, for protoplanets with lower masses the boundaries between CPDs and protoplanetary gas are blurred since the pressure support becomes increasingly important compared to the gravity of protoplanets. As a result, the structure of CPD formed around 4M$_{\Earth}$ protoplanet looks more like an oblate spheroid.}

\begin{figure}
\epsscale{1.1}
\plotone{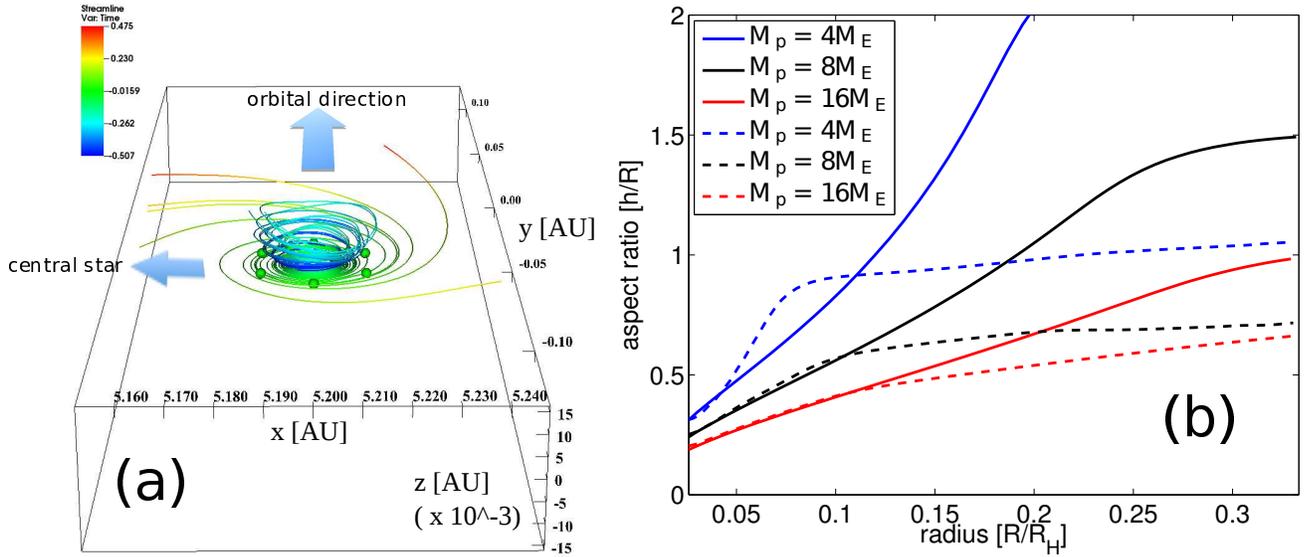}
\caption{(a) The bird's-eye view of streamlines (solid-lines) traced in a CPD formed around a protoplanet of 16M$_{\Earth}$.  The dots represent the zero-time points for tracing the streamlines.  They are arranged on a circle with its normal along the $z-$axis  and centered at $(r,\phi,z)=(5.2{\rm AU}, 0, 0.001{\rm AU})$. The radius of the circle is chosen to be 0.016AU defined as the size of the CPD (cf. Tabel~\ref{table:properties}). These streamlines are traced both forward (redder, positive number) and backward (bluer, negative number) using colors to denote the time elapse from the zero-time points.  The unit of time is $({\rm AU}/{\rm km})\cdot s$. (b) The solid-lines denote the estimated scale heights of CPDs using semi-analytic models for planets with 4, 8 and 16M$_{\Earth}$, while the dashed-lines are the corresponding scale heights \hhwang{obtained by azimuthally averaging the structure of CPDs over the last 8 orbital times of simulations}.  \label{fig:scaleheight_streamline}}
\end{figure}

\hhwang{The CPD formed around the protoplanet of 16M$_{\Earth}$ reaches a steady state in a few orbital times, while} those surrounding protoplanets of 4 and 8M$_{\Earth}$ develop non-steady spiral shocks and remain constantly disturbed by protoplanetary gas. \hhwang{This phenomenon can be understood by looking at the flow lines of protoplanetary gas (Fig.~\ref{fig:scaleheight_streamline}a) together with the density contrast between CPDs and the surrounding protoplanetary gas (Fig.~\ref{fig:edge_face_view}). Figure~\ref{fig:scaleheight_streamline}a shows the bird's-eye view of six streamlines that follow the tracks of protoplanetary gas moving through the CPD formed around a protoplanet of 16M$_{\Earth}$. These streamlines are integrated both forward (positive) and backward (negative) from the zero-time points shown as the green dots. The time elapse from the zero-time points is coded with different colors along the streamlines. That is, the bluer color is more backward in time while the redder more forward in time. The green dots are uniformly placed on a circle centred at $(r,\phi,z)=(5.2,0,0.001)$AU with its normal along the $z$-axis. The radius of the circle is chosen to be 0.016AU defined as the size of the CPD using the turning point seen in the plot of specific angular momentum (discuss below). The streamlines show that high altitude protoplanetary gas accretes onto the CPD vertically, circling the central protoplanet for several times before it return as again the protoplanetary gas. This figure implies that this CPD is constantly exchanging angular momentum and material with protoplanetary gas and explains how the CPD get disturbed. Nevertheless, the high density contrast as seen from the edge-on view for the 16M$_{\Earth}$ protoplanet (Fig.~\ref{fig:edge_face_view}c) makes this CPD less subject to the disturbance from the protoplanetary gas. On the contrary, those CPDs formed around 4 or 8M$_{\Earth}$ protoplanets develop non-steady shocks due to the impinging of high altitude protoplanetary gas.}

\hhwang{Another important conclusion we can draw from Fig.~\ref{fig:scaleheight_streamline}a is that for sub-Neptune-mass protoplanets \wang{a fraction of} gas situated inside \wang{but near} $R_{\rm H}/10$ is not bound to the central protoplanets. This naturally leads to the asymmetry of CPD as seen in Fig.~\ref{fig:edge_face_view}c, since the protoplanetary gas in the vicinity of protoplanet is also not entirely symmetric with respect to the protoplanet. The lopsidedness of CPD is sustained by the protoplanetary gas and may have an impact on the long-term planet migration due to its proximity.}

\hhwang{The scale height of an isothermal CPD can be estimated analytically by assuming that the CPD is vertically hydrostatic}, i.e.,
\begin{equation}
-\frac{c^2_s}{\rho}\frac{\partial \rho}{\partial z} = \frac{\partial \Phi}{\partial z}. 
\end{equation}
As a result, $\rho(R,z)=\rho_0(R)\exp[-(\Phi(R,z)-\Phi(R,z=0))/c^2_s]$, where the total potential $\Phi(R,z)$ is defined by Eq.~(5) and $\rho_0(R)$ is the volume density in the midplane. For a given \hhwang{planetocentric radius}, the scale height, $h$, is then estimated as the standard deviation of \hhwang{a Gaussian function fitted for the vertical volume density}. 

The estimated aspect ratios of CPDs as functions of \hhwang{planetocentric distance} for our models are shown as the solid curves in Fig.~\ref{fig:scaleheight_streamline}b. The radial distances are normalized with the corresponding Hill radius. Evidently, isothermal CPDs are flaring with increasing distance. The degree of disk flaring increases with decreasing protoplanetary mass. This plot indicates that \hhwang{disk-like} objects \hhwang{are expected to} form around sub-Neptune-mass protoplanets and should be resolved with sufficient spatial resolution. Based on this result, if one defines a radius that corresponds to the aspect ratio $h/R = 0.5$ as the size of a disk, with our fineset spatial resolution, the inner part of CPDs with aspect ratios less than 0.5 will be covered with 7, 14 and 29 cells for protoplanets with masses 4, 8 and 16M$_{\Earth}$, respectively. The radii that correspond to the aspect ratio $h/R=0.5$ can be either read directly from Fig.~\ref{fig:scaleheight_streamline}b or found in Table~\ref{table:properties} \hhwang{(column 4)}.  The choice of $h/R=0.5$ is somewhat arbitrary. However it helps us estimate the spatial resolution needed to resolve a CPD. \hhwang{We note that} our spatial resolution \hhwang{placed around} protoplanets is fairly sufficient to resolve a CPD formed around \hhwang{the 16M$_{\Earth}$} protoplanet, while it only marginally resolves the CPD formed around \hhwang{the 4M$_{\Earth}$} protoplanet.

The dashed curves shown in Fig.~\ref{fig:scaleheight_streamline}b are the aspect ratios of CPDs extracted from our numerical simulations \hhwang{using the Gaussian fit. These curves are obtained by azimuthally averaging the structure of CPDs over the last 8 orbital periods of simulations.} If again taking \hhwang{a radius that corresponds to $h/R=0.5$ as the disk size (see Table~\ref{table:properties}, column 5), the results obtained for the models of 8 and 16M$_{\Earth}$ closely follow the corresponding solid lines until 0.1$R_{\rm H}$, which is consistent with disk size defined by the turning point of specific angular momentum. The deviation seen in the model of 4M$_{\Earth}$ is expected from the edge-on view shown in Fig.~\ref{fig:edge_face_view}a, since the vertical extent of the CPD is blurred with the surrounding protoplanetary gas. Although the gas situated beyond 0.1$R_{\rm H}$ should not be considered as part of CPDs, the aspect ratios obtained for all simulation models are generally thinner than the corresponding solid lines. This is because, within the Hill radius, protoplanetary gas is vertically perturbed by the gravity of the protoplanet, as a result, ram pressure exerted from the top may significantly reduce the expected scale heights.}

\begin{figure}
\epsscale{1.1}
\plotone{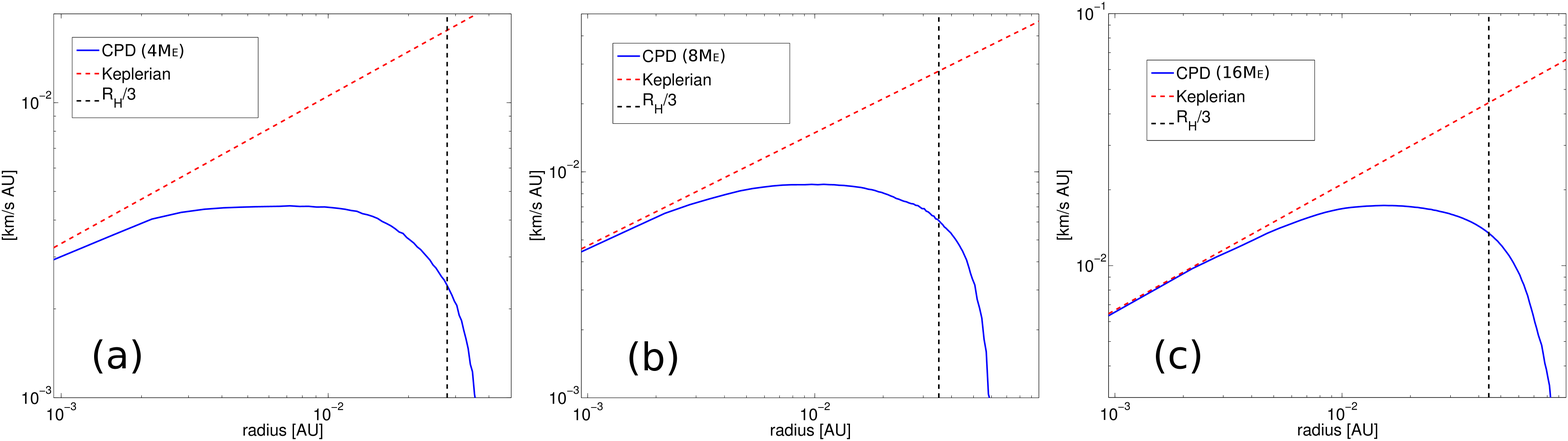}
\caption{Specific angular momenta measured in the rotating frame as functions of planetocentric radius for planets with masses of 4, 8 and 16M$_{\Earth}$, respectively. The vertical black-dashed lines indicate the corresponding one-\hhwang{third} of Hill radius. The red-dashed lines denote the ideal specific angular momenta for Keplerian disks. \label{fig:angmom}}
\end{figure}

Figure~\ref{fig:angmom} shows the specific angular momentum measured in the rotating frame as a function of \hhwang{planetocentric} radius. The red-dashed lines represent the ideal Keplerian motions, while the vertical black-lines mark the locations of one-third Hill radii. The inner CPD surrounding \hhwang{the 16M$_{\Earth}$ protoplanet} is nearly Keplerian, while the one surrounding 4M$_{\Earth}$ is sub-Keplerian. This result is consistent with the edge-on view shown in Fig.~\ref{fig:edge_face_view}, since the latter CPD is embedded in a partially pressure-supported envelope. If \hhwang{taking} the turning point seen in the specific angular momentum as the size of CPDs \citep{Ayl09b,Bu13},  as \hhwang{listed} in Table~\ref{table:properties}, the disk sizes will be roughly \hhwang{$R_{\rm H}/10$}, much less than $R_{\rm H}/3$ suggested by the kinematic argument \citep{Qui98}. 

\section{Discussion and Summary}

\subsection{Discussion}

\hhwang{In the isothermal limit, the gas behaviour around sub-Neptune-mass protoplanets in this work is different from that of \citet{Ayl09a}. While they found no disk-like structure around protoplanets of mass less than 33$M_{\Earth}$, our results indicate a CPD of expected vertical profile can reasonably form around a protoplanet with mass at least down to 8$M_{\Earth}$. The exact reason that leads to the discrepancy is not clear. It seems that the difference in spatial resolution should not be the main cause, since those adopted in \citet{Ayl09a} ($\approx 3\times 10^{-4}R_{\rm H}$) are much better than what is done in this work ($\approx 5\times 10^{-3}R_{\rm H}$ for the 16$M_{\Earth}$ protoplanet). Beyond this specific aspect, some differences between \citet{Ayl09a} and our work still exist. Gas viscosity, which presumably originates from magnetic turbulence, is not explicitly included in this work. On the contrary, artificial viscosity is a standard procedure when using a particle-based hydrodynamic code. The impact of viscosity on the formation and on the structure of CPDs is not well understood. Two-dimensional simulations in shearing boxes by \citet{Bu13} suggested that viscosity may be responsible for transferring angular momentum out of disks and facilitates mass accretion onto CPDs. On the other hand, as shown in this work, the gas motion around low-mass protoplanets is fully three-dimensional. It is not clear how viscosity would work as a CPD accretes gas vertically. It has been tested and reported that a grid-based code is especially suitable for problems in which the physics of interest is in the region of rapidly changing density \citep{Tas08}. As shown in Figure~\ref{fig:edge_face_view}c, the density contrasts can be more than three orders of magnitude between CPDs and the surrounding gas. Whether or not differences in the numerical schemes can generate the main differences requires further investigation.}

\hhwang{In this work, the sizes of CPDs determined by the turning points of specific angular momenta are substantially smaller than $R_{\rm H}/3$, which is obtained from a pure kinematic argument \citep{Qui98}. The size of $R_{\rm H}/3$ is usually expected for the CPDs around more massive protoplanets \citep{Ayl09b}. The turning point of the specific angular momentum marks a radius beyond which gas does not orbit the protoplanet, and is often taken as the outer edge of a CPD \citep{Ayl09b,Bu13}. On the other hand, the kinematic argument neglects the effect of thermal pressure and is applied to protoplanets massive enough such that their gravitational forces dominate over the thermal pressure inside the Hill radii. In this case, the sizes of CPDs are limited by the conservation of angular momenta of the inflow gas. For the low-mass regime explored in this work, thermal pressure is important compared with the gravitational forces of the protoplanets inside the Hill radii. Keplarian disks can only be expected for the regions close enough to the protoplanets (due to the inverse square law of gravity), resulting in smaller CPDs around low-mass protoplanets. The second and the third columns of Table~\ref{table:properties} indicate that more massive protoplanets tend to have larger CPDs in terms of Hill radii. We conclude that the size of CPDs increases with the mass of protoplanets and $R_{\rm H}/3$ should be taken as an upper limit of disk size.}

\hhwang{The sizes of CPDs may impact calculations of torques exerted on protoplanets. CPD plus its protoplanet are often assumed to be a gravitational bound system so that their internal interaction would not have long-term effect on the protoplanetary migration. Based on this assumption, either a planetocentric radius inside which the interaction between the CPD and the protoplanet is excluded or a softening length comparable to the size of CPDs is often introduced in the study of planet migration. In fact, the region that can be excluded in the calculation of migration causing torques is still lack of consensus, especially for low-mass protoplanets, primarily due to the poor spatial resolution in the vicinity of protoplanets. Efforts have been put to quantify the region to exclude the migration causing torques for massive protoplanets. Two-dimensional simulations suggest that a significant fraction of the total torque exerted on a massive protoplanet is from the region $R_{\rm H}/2<R<R_{\rm H}$ \citep{Cri09}.  Three-dimensional simulations by \citet{Ayl09b} showed that a CPD surrounding a Jovian mass protoplanet may extend to $\approx R_{\rm H}/3$ inside which the gas exerts no migration causing torque to the protoplanet. \citet{Ayl10} explored thermal effects on the Type I migration timescale for low-mass protoplanets ($10-33M_{\Earth}$) using more realistic protoplanetary surfaces located at 0.03$R_{\rm H}$, where the gravitational forces of protoplanets diminish to zero. The treatment of protoplanetary surface allows the envelope close to the core to develop self-consistently. Since in this mass regime, there exists discrepancy between their and our results, a straightforward interpretation can not be drawn. More recently, \citet{Tan12} studied the gas accretion flow onto a CPD formed around a Jovian mass protoplanet using three-dimensional local nested-grid hydrodynamic simulations with a spatial resolution one-fourth of the present Jupiter radius. They found that the outward radial velocity increases significantly at a planetocentric distance about $R_{\rm H}/5$, pushing the planetocentric boundary, within which one may exclude gaseous torque, further inside $R_{\rm H}/5$. On the other hand, our results indicate that the torques exerted on low-mass protoplanets require a careful calculation for the gas situated inside one-tenth of Hill radius. Owing to the proximity of CPDs to the protoplanets, the lopsidedness of CPD sustained by protoplanetary gas moving in and out of the CPD may affect planet migration in the long run. The softening length should be used with caution when a low-mass planet is free to migrate.} 

\hhwang{Since protoplanetary gas moves in and out of the CPDs, the systems around low-mass protoplanets may be easily perturbed. As shown in Figure~\ref{fig:edge_face_view}a,b, nonsteady spiral shocks may appear in CPDs surrounding protoplanets with masses less than 16$M_{\Earth}$. These nonsteady structures are not likely due to numerical instability, since they are only prominent in the models with lower masses, i.e., 4 and 8$M_{\Earth}$, and disappear in the model with 16$M_{\Earth}$. This trend is physically expected since lower density contrast between the CPDs and gas infalling at high altitudes as seen in Figure~\ref{fig:edge_face_view}a,b makes these CPDs more vulnerable to disturbances from protoplanetary gas. We also performed two-dimensional local and global simulations (unpublished) for the same mass regime using the same numerical code, Antares. For the two-dimensional global simulations, CPDs get disturbed with epicyclic frequency, though the disturbances is from the midplane rather than from high altitudes. By contrast, only steady CPDs are observed for two-dimensional local simulations, where the outgoing/injecting boundary conditions are applied in the azimuthal direction \citep{Tan02,Bu13}. This simple tests suggest that the disturbance is a feedback from the perturbed protoplanetary gas, which can only be properly modeled in global simulations. The nonsteady shocks are more likely due to physical origin rather than a numerical artifact since we would otherwise expect to see them as well in two-dimensional local simulations. }

\hhwang{The isothermal three-dimensional global simulations in this work show interesting implications for planet formation, without the inclusion of viscosity, grains, self-gravity and planet migration. Current understanding of the core accretion model was mainly based on results obtained in one-dimensional calculations \citep{Pol96,Lis09,Mov10,Mor12}, which adopted the assumed spherical symmetry of core-nucleated gas envelopes surrounding solid cores of 1-15 earth masses. \citet{Lis09} and \citet{Mov10} suggest, due to the grain growth, the grain opacities in the envelopes of protoplanets can be three to four orders of magnitude less than the interstellar level, resulting in more rapid heat loss. Our results indicate that in the limit of vanishing opacity, instead of direct contraction, centrifugal barrier in the conservation of angular momentum leads to the formation of CPD. As a result, low grain opacities may imply that one-dimensional models of the protoplanetary envelopes might be geometrically oversimplified. However, when thermal support is important, the protoplanets may remain approximately spherical up to 100$M_{\Earth}$ then undergo gravitational collapses that lead to the formation of CPDs \citep{Ayl12,Lis09,Mor12}. }

\subsection{Summary}
\hhwang{We address the formation of CPDs around sub-Neptune-mass protoplanets using three-dimensional global hydrodynamic simulations. The CPDs, inside which satellites grow, are believed to form after the protoplanets have accreted most of their final masses. While inferences from observations expect the CPDs to be cool and optically thin, the processes of forming CPDs and accreting material are not well understood. By implementing nested-mesh-refinement to our Antares code, protoplanetary disks and CPDs can be properly resolved and evolved together. Cylindrical coordinates are adopted for the effects of curvature. An initial condition in equilibrium was carefully prepared to study the interaction between the protoplanetary and circumplanetary disks. Our findings are summarized as follows:

\begin{itemize}
 \item The size of isothermal CPDs formed around sub-Neptune-mass protoplanets is estimated to be $R_{\rm H}$/10 using the turning point of specific angular momentum. 
 \item The vertical structures of CPDs formed around 8 and 16$M_{\Earth}$ protoplanets fit well with the estimated scale heights based on the assumption of vertical hydrostatic equilibrium, while the CPD around 4$M_{\Earth}$ resembles an oblate spheroid embedded in protoplanetary gas. 
 \item The streamlines around the protoplanets enter the CPDs almost vertically from high altitudes and return to the protoplanetary disk in heliocentric orbits near the midplane. A CPD is an open system, constantly exchanging angular momentum and material with the outer protoplanetary disk. The lopsideness of CPDs with respect to the protoplanet is maintained by flows in and out of the system for tens of \wang{orbits}, whose symmetry is broken by geometric curvature. This indicates that torque contributed by gas inside $R_{\rm H}$/10 might have a long-term influence on planet migration. 
 
 \item The CPDs formed around 4 and 8$M_{\Earth}$ protoplanets may develop non-steady shocks due to disturbances in the protoplanetary gas, while those CPDs around more massive protoplanets are relatively steady because of the high density contrast between CPDs and the ambient gas. 
 \item In the limit of vanishing opacity, the presence of a CPD implies that a centrifugal barrier needs to be overcome before a low-mass protoplanet can grow to a Jupiter size. 
\end{itemize}
}

\acknowledgments
\hhwang{The authors would like to acknowledge the support of the Theoretical Institute for Advanced Research in Astrophysics (TIARA) based in Academia Sinica Institute of Astronomy and Astrophysics (ASIAA). D. F. Bu was supported by the Natural Science Foundation of China (grants 11103059). The authors thank the referee for comments which helped to improve the clarity and presentation of this paper. Thanks to Mr. Sam Tseng for assistance on the computational facilities and resources (TIARA cluster).}

 \bibliography{CPDbib}

\clearpage
\begin{table}
\begin{center}
\caption{Properties of CPDs \label{table:properties}}
\begin{tabular}{ccccccc}
\tableline\tableline 
planet mass & \hhwang{$R_{H}$} & disk size \tablenotemark{a} & radius of $h/R=$0.5   &  radius of $h/R=$0.5 & \hhwang{$R_{\rm core}$\tablenotemark{b,c}} & \hhwang{$N_{R_{H}}$\tablenotemark{d}}\\
\hhwang{$[{\rm M}_{\Earth}]$} &    [AU]        &   [AU]       &  [$R_H$](estimated) &  [$R_H$](simulation) & \hhwang{[AU]} & \\
\tableline
\hhwang{4} & 0.0837 &   0.008 &    0.055 & \hhwang{0.049} & \hhwang{$9.0\times 10^{-5}$}   & \hhwang{134}\\
\hhwang{8}  & 0.1054 &    0.01 &    0.085 & \hhwang{0.082} & \hhwang{$1.13\times 10^{-4}$}  & \hhwang{168}\\
\hhwang{16}   & 0.1328 &   0.016 &    0.137 & \hhwang{0.164}  & \hhwang{$1.43\times 10^{-4}$} & \hhwang{212}\\
\tableline
\end{tabular}
\end{center}
\tablenotetext{a}{\hhwang{The disk size is determined by the turning point of specific angular momentum shown in Fig.~\ref{fig:angmom}.}}
\tablenotetext{b}{\hhwang{The numerical resolution near the protoplanets is $6.25\times 10^{-4}$AU.}}
\tablenotetext{c}{\hhwang{The mean density of the ice-rock core adopted here is 2.35 g cm$^{-3}$, which is estimated based on the model for the core of Uranus \citep{Pod95}.}}
\tablenotetext{d}{\hhwang{The effective number of cells used to resolve a Hill radius in terms of the finest spatial resolution.}}
\end{table}

\end{document}